\documentclass{elsart}

\usepackage[square,comma]{natbib}
\usepackage{graphicx}
\usepackage{pxfonts}
\usepackage{lineno}

\usepackage{amssymb}
\usepackage{footnote}
\usepackage{multirow}
\usepackage{varwidth}
\usepackage{array}
\usepackage{epstopdf}
\usepackage{url}

\journal{}

\begin{document}

\thispagestyle{empty}
\begin{Large}
\textbf{DEUTSCHES ELEKTRONEN-SYNCHROTRON}

\textbf{\large{Ein Forschungszentrum der
Helmholtz-Gemeinschaft}\\}
\end{Large}

DESY 17-047

April 2017

\begin{eqnarray}
\nonumber &&\cr \nonumber && \cr \nonumber &&\cr
\end{eqnarray}
\begin{eqnarray}
\nonumber
\end{eqnarray}
\begin{center}
\begin{Large}
\textbf{Radiation by Moving Charges}
\end{Large}
\begin{eqnarray}
\nonumber &&\cr \nonumber && \cr
\end{eqnarray}

\begin{large}
Gianluca Geloni,
\end{large}
\textsl{\\European XFEL GmbH, Hamburg}
\begin{large}

Vitali Kocharyan and Evgeni Saldin
\end{large}
\textsl{\\Deutsches Elektronen-Synchrotron DESY, Hamburg}
\begin{eqnarray}
\nonumber
\end{eqnarray}
\begin{eqnarray}
\nonumber
\end{eqnarray}
ISSN 0418-9833
\begin{eqnarray}
\nonumber
\end{eqnarray}
\begin{large}
\textbf{NOTKESTRASSE 85 - 22607 HAMBURG}
\end{large}
\end{center}
\clearpage
\newpage

\begin{frontmatter}



\title{Radiation by Moving Charges}


\author[XFEL]{Gianluca Geloni,}
\author[DESY]{Vitali Kocharyan,}
\author[DESY]{Evgeni Saldin}
\address[XFEL]{European XFEL GmbH, Hamburg, Germany}
\address[DESY]{Deutsches Elektronen-Synchrotron (DESY), Hamburg, Germany}

\begin{abstract}

It is generally accepted that in order to describe the dynamics of relativistic particles in the laboratory (lab) frame it is sufficient to take into account the relativistic dependence of the particles momenta on the velocity. This solution of the dynamics problem in the lab frame makes no reference to Lorentz transformations. For this reason they are not discussed in particle tracking calculations in accelerator and plasma physics. It is generally believed  that the electrodynamics problem can be treated within the same "single inertial frame" description without reference to Lorentz transformations. In particular, in order to evaluate  radiation fields arising from charged particles  in motion  we need to know their  velocities  and  positions as a function of the lab frame time $t$. The relativistic motion of a particle in the lab frame is described by Newton's second law "corrected"  for the relativistic dependence of momentum on velocity. It is assumed in all standard derivations that one can perform identification of the trajectories in the source part of the usual Maxwell's equations with the  trajectories  $\vec{x}(t)$ measured (or calculated by using the corrected Newton's second law) in the lab frame. This way of coupling fields and particles is considered since more than a century as the relativistically correct procedure. We argue that this procedure needs to be changed, and we demonstrate the following, completely counterintuitive statement: the results of conventional theory of radiation by relativistically moving charges are not consistent with the principle of relativity. In order to find the trajectory of a particle in the lab frame consistent with the usual Maxwell's equations,  one  needs to solve the dynamics equation in manifestly covariant form by using the coordinate-independent proper time $\tau$ to parameterize the particle world-line in space-time. We show that there is a difference between  "true" particle trajectory $\vec{x}(t)$  calculated or measured in the conventional way, and covariant particle trajectory $\vec{x}_{cov}(t)$ calculated by projecting the  world line  to the lab frame  ($t(\tau), x_1(\tau), x_2(\tau), x_3(\tau)$) and using the lab time $t$ to parameterize the trajectory curve. In other words, for a relativistic motion accelerated along a curved trajectory, the results of conventional particle tracking differ from those of covariant particle tracking. The difference is only due to a choice of convention, but only $\vec{x}_{cov}(t)$ is consistent with the usual Maxwell's equations. This essential point has never received attention in the physical community. As a result, a correction of the conventional synchrotron-cyclotron radiation theory is required.
\end{abstract}

%
%

%
\end{frontmatter}



\section{ Introduction }

The special theory of relativity is based on a space-time with pseudo-Euclidean geometry. The principle of relativity can in fact be seen as a  consequence of the space-time geometry. To simplify computations one often works with the components of vectors and tensors rather than with geometric objects. Each Lorentz frame (i.e. inertial frame with a Lorentz system of coordinates) gives a coordinate dependent representation of any geometric object or relation, and transformation from one Lorentz frame to another are sometimes needed. These coordinate transformations take inertial coordinate system onto inertial coordinate system and keep the form of laws of nature.

It is generally accepted that in order to describe the dynamics of relativistic particles in a single inertial  frame, there is no need to use the laws of relativistic kinematics. It is sufficient to take into account the relativistic dependence of the particles momenta on the velocity. With this correction factor properly taken into account,  Newton's laws  are in full agreement with the principle of relativity. Theoretical treatment of relativistic particle dynamics in the laboratory (lab) reference frame involves only corrected Newton's second law.  Note that this solution of the dynamics problem in the lab frame makes no reference  to Lorentz transformations, and this is a reason why they are not discussed in particle tracking calculations involved in accelerator and plasma physics.

It is generally believed that the electrodynamics problem can be treated within the same "single inertial frame" description without reference to Lorentz transformations (see standard textbooks, e.g. \cite{LL,J}). When the lab  frame is chosen with a Lorentz system of coordinates in it, then  the usual Maxwell's equations apply, and all results of classical electromagnetism are recovered in that frame.  It is assumed in all standard derivations that usual Maxwell's equations and corrected Newton's second law can explain all experiments that are performed in a single inertial frame, for instance the lab frame. In particular, in order to evaluate the radiation fields arising from a collection of sources  we only need to specify  velocities  and  positions of the charged particles involved  as a function of the lab frame time $t$. In its turn, the relativistic motion of these particles in the lab frame is described by the corrected Newton's second law. This coupling of Maxwell's equations and the corrected Newton's equation  is  commonly accepted as useful method in accelerator and plasma physics and, in particular, in analytical and numerical calculations of synchrotron and cyclotron radiation properties (see e.g. \cite{DU,Eck}). Such approach to relativistic dynamics and electrodynamics usually forces the physicist to believe that the description of  radiation by a relativistic charged particle is possible without detailed knowledge of the theory of relativity.

In our previous publications \cite{OURS1,OURS2,OURS3, OURS4} we argued that  there is a fundamental, surprising disagreement between  this commonly accepted way of coupling electrodynamics and dynamics in the lab frame  and the principle of relativity.  In particular, we argued that it  leads to a strong qualitative disagreement between conventional results of synchrotron-cyclotron radiation theory and experiments. In other words, we support the following counterintuitive statement: in the lab reference frame there is incompatibility between, on the one hand, the trajectory of a charged particle $\vec{x}(t)$ calculated by using the corrected Newton's second law and, on the other hand, the usual Maxwell's equations.

This incompatibility is only due to an inconsistent choice of conventions used in the solution of dynamics and electrodynamics problem. In order to find the particle trajectory compatible with the usual  Maxwell's equations in the lab frame, one needs to solve the dynamics equation in manifestly covariant form by using the coordinate-independent proper time $\tau$ to parameterize the world-line that describes the particle evolution. As we will see, the trajectory $\vec{x}_{cov}(t)$, which is compatible with the usual Maxwell's equations can be found by using the manifestly covariant dynamics equation. From  the lab frame it is viewed as the result of successive  Lorentz transformations between the  lab frame with Lorentz coordinates and the co-moving sequence of Lorentz reference frames. The composition law that follows from group properties of Lorentz transformations is used to express the conditions of co-moving sequence of frames tracking a particle. The motion of a particle in the lab frame coordinates is included in the functions ($t(\tau), x_1(\tau), x_2(\tau), x_3(\tau)$), which basically are, at any $\tau$, components of a four-vector describing an event in a space-time. Thus, if one  chooses the lab time $t$ as a parameter for the trajectory curve, after inverting the relation $t = t(\tau)$ the space position vector  of a particle in the lab frame can  be assumed to have the functional form $\vec{x}_{cov}(t)$. We will see that there is a difference between conventional particle trajectory $\vec{x}(t)$, calculated by solving the corrected  Newton's equations, and the covariant particle trajectory $\vec{x}_{cov}(t)$, calculated by projecting the world line  onto the lab frame. This essential point has never received attention in the physical community. As a result, the difference between conventional particle trajectory  and covariant particle trajectory seems to have been almost entirely overlooked by many physicists including J. Schwinger \cite{SCH}, who developed results of synchrotron radiation theory by using the usual Maxwell's equations  and  $\vec{x}(t)$, instead of $\vec{x}_{cov}(t)$.

The statement that there is difference between the two trajectories  $\vec{x}(t)$ and $\vec{x}_{cov}(t)$ does not mean that the conventional trajectory  $\vec{x}(t)$ is incorrect.
Within the frame of dynamics only,  both trajectories describe correctly the same physical reality. Different expressions for the particle trajectories are different only because they are based on the use of different clocks synchronization conventions. Whenever we have a theory containing an arbitrary convention, we should examine what parts of the theory depend on the choice of that convention and what parts do not. We may call the former convention-dependent, and the latter convention-invariant parts. Clearly, physically meaningful  results must be convention-invariant.  We state that  the difference between the two trajectories  $\vec{x}(t)$ and $\vec{x}_{cov}(t)$ is convention-dependent and has no direct physical meaning. In fact, once more, different expressions for the particle trajectory in a single (e.g. lab) reference system arise from the use of different synchronization conventions, and different types of clocks synchronization simply provide different time coordinates that describe the same reality. In particular, we cannot specify any experimental method by which simultaneity between two events in different places can be decided. In other words, the determination of simultaneous events imply the choice of a convention. In a similar fashion,  in order to measure the speed of a particle, one first has to synchronize the clocks that measure the time interval as the particle travels between two given points in space. Therefore it can be said that, consistently with conventionality of simultaneity, the value of the particle speed is also a matter of convention and has no definite objective meaning.

The conventional nature of distant simultaneity in special relativity is not to be confused with the relativity of simultaneity. Clearly, the conventionality of simultaneity within a single inertial frame is quite distinct from the relativity of simultaneity. The theory of relativity showed that it is meaningful to discuss about the simultaneity of events only relatively to a given reference frame. However, it should be clear that even within a single reference frame the definition of simultaneity of events  is matter of convention \cite{Frid}.

The usual non-covariant approach leading to the calculation (or the measurement) of the particle trajectory  $\vec{x}(t)$ in the lab frame is understood  to deal not with geometric, tensor quantities in 4D space-time, but rather with quantities from  the "3+1" space and time. In other words, in the lab frame, Minkowski space-time "splits up" into three-dimensional space and one-dimensional time, yielding a separate "3+1" space and time. Within the lab frame, this looks precisely the same as in Newtonian kinematics: there is an Euclidean three-space, a global time $t$, and the Newton's second law of motion, albeit corrected for the relativistic dependence of momentum on velocity \cite{Frid}.  In the lab "3+1" space and time we cannot associate the trajectory of a charged particle with $\vec{x}_{cov}(t)$. For instance, a trajectory measurement would give the usual noncovariant result $\vec{x}(t)$.  In this sense, in the lab frame the function  $\vec{x}_{cov}(t)$  has only a formal significance. The covariant trajectory must be used only because we need to solve electrodynamics problem based on the use of Maxwell's equations in their usual form.

One might choose to use the noncovariant trajectory $\vec{x}(t)$, but the price to pay would be a change in the form of Maxwell's equations. In fact, later on we will see that
there are two satisfying ways of coupling fields and particles. The first, Einstein's way, consists in using relativistic kinematics for the description of the particle evolution and the usual Maxwell's equations. The second  way consists, instead, in using the noncovariant  trajectory $\vec{x}(t)$ together with a "translation" of Maxwell's equations to the "3+1" space and time  convention, which we will call everywhere in this paper, the "absolute time convention". We will show that the transformation of Maxwell's equations to the absolute time convention can be found by means of the metric tensor $g_{\mu\nu}$, which has in this case non-diagonal components.

If one chooses the second way, one is faced with the problem of solving a modified version of Maxwell's equations:  in the "3+1" space and time we have much more complicated field equations to deal with. To get around this difficulty, we observe that these equations  can always be simplified. One only needs to make a change of time and spatial variables. Transforming to new variables leads to the usual Maxwell's equations.  However, in these new variables the non-covariant trajectory $x(t)$ is automatically transformed to the covariant trajectory $\vec{x}_{cov}(t)$ that is the trajectory  compatible, as we discussed above,  with the usual Maxwell's equations.  The bottom line is that  the change of variables in the "3+1" space and time is nothing but a Lorentz transformation, which  is now, however,  only  used as a mathematical device yielding the solution of the electrodynamics problem with minimal effort.

To summarize our arguments, we can say that the usual way of coupling fields and particles in the "3+1" space and time has been considered for more than a century as the relativistically correct procedure to follow. Textbooks on electrodynamics describe how the properties of radiation by relativistic moving charges can be calculated by using the lab frame, without any reference to Lorentz transformations,  and coupling the usual Maxwell's equations with the corrected  Newton's second law that yield  the trajectory $\vec{x}(t)$.

At variance, in  sections 2, 3 and 4 of this paper we explicitly challenge this procedure, showing that the usual Maxwell's equations should be coupled instead to the covariant trajectory $\vec{x}_{cov}(t)$. In those sections we also present a practical case study for illustrating  the difference between conventional and covariant trajectories. In section 5 we discuss the aberration of light effect as an example illustrating the electrodynamics of moving charges. We have chosen this particular example because it is relatively simple, although it can be seen as a prototype that can be generalized for the description of all radiation phenomena. In section 6  we analyze our theory in connection with experiments involving relativistic electrons and, finally, in section 7 we come to conclusions.

\section{Error in coupling fields and particles  within a single inertial frame}

As discussed in the introduction, it is generally accepted that in order to describe the dynamics of relativistic particles in the lab reference frame, which we assume inertial, one only needs to take into account the relativistic dependence of the particles momenta on the velocity. In other words, the treatment of relativistic particle dynamics involves only a corrected Newton's second law. In a fixed lab frame, we consider an electric field $\vec{E}$ and a magnetic field $\vec{B}$. They interact with a charged particle in accordance with

\begin{eqnarray}
&& \frac{d\vec{p}}{dt} = e\left(\vec{E} + \frac{\vec{v}}{c}\times \vec{B}\right) ~,\cr &&
\vec{p} = m\vec{v}\left(1 - \frac{v^2}{c^2}\right)^{-1/2}~ ,\label{N}
\end{eqnarray}
where the particle rest mass, charge, and velocity are denoted by $m$, $e$, and $\vec{v}$  respectively. The Lorentz force law, plus measurements on the components of the acceleration of test particles, can be viewed as defining the components of electric and magnetic fields. Once the field components are found in this way, they can be used to predict the accelerations of other particles \cite{MTW}.

Thus, within a single inertial lab frame there is an Euclidean three-space, a global, absolute time $t$, and the corrected Newton's second law of motion. Notice that the solution of the dynamics problem  in the lab frame does not involve any Lorentz transformation. In other words, the laws of relativistic dynamics within the lab frame, where Minkowski space-time "splits up" into three-dimensional space and one-dimensional time,  is very simple: aside for a straightforward correction, it looks precisely the same as in Newtonian dynamics.

Going to the electrodynamics problem, the differential form of  Maxwell's equations describing electromagnetic phenomena in the same inertial lab frame (in cgs units)  is given by the following expressions:

\begin{eqnarray}
&& \vec{\nabla}\cdot \vec{E} = 4 \pi \rho~, \cr && \vec{\nabla}\cdot
\vec{B} = 0~ , \cr && \vec{\nabla}\times \vec{E} =
-\frac{1}{c}\frac{\partial \vec{B}}{ \partial t}~,\cr &&
\vec{\nabla}\times \vec{B} = \frac{4\pi}{c}
\vec{j}+\frac{1}{c}\frac{\partial \vec{E}}{\partial t}~. \label{Max}
\end{eqnarray}
Here the charge density $\rho$ and current density $\vec{j}$ are written as

\begin{eqnarray}
&&\rho(\vec{x}, t) = \sum_{n} e_n\delta(\vec{x} - \vec{x}_n(t)) ~,\cr &&
\vec{j}(\vec{x},t) = \sum_{n} e_n\vec{v}_n(t)\delta(\vec{x} - \vec{x}_n(t))~, \label{CD}
\end{eqnarray}

where $\delta(\vec{x} - \vec{x}_n(t))$ is the three-dimensional Dirac $\delta$-function, while $m_n, e_n, \vec{x}_n(t)$, and $\vec{v}_n =   d\vec{x}_n(t)/dt$ denote respectively rest mass, charge, position, and  velocity of the $n$-th particle involved in the electrodynamics process. To evaluate  radiation fields arising from the external sources in Eq. (\ref{CD}), we need to know the velocity $\vec{v}_n$ and the position $\vec{x}_n$ as a function of the lab frame time $t$. As discussed above, it is generally accepted  that the equations of motion, which describe how the coordinates of the charged particles change with the lab time $t$, are described by the corrected Newton's second law Eq. (\ref{N}).

In  previous work \cite{OURS1,OURS2,OURS3, OURS4} we argued that this way of coupling fields and particles in the lab frame, which is considered in all standard treatments, is surprisingly incorrect. We pointed out that only a solution of dynamics and electrodynamics equations in manifestly covariant form results in the correct coupling between fields  and particles. In the next sections we explain our reasons in support to the following, completely counterintuitive statement: in the relativistic limit, the usual algorithm  for solving  Maxwell's equations in the lab frame, i.e. with current and charge density created by  particles moving along the trajectories $\vec{x}_n(t)$, described by Eq. (\ref{N}), is at odds  with the principle of relativity.

\section{Dynamics. Space-time geometry}

A geometrical view of physics often yields great conceptual clarity. Space and time form a unique four-dimensional continuum of events with pseudo-Euclidean geometry. This is the essence of the special theory of relativity, and has consequences on all physical phenomena. In particular, all physical laws expressed in terms of geometric objects automatically include the principle of relativity.  Physical laws, and in particular the dynamics equations,  can be expressed as tensor equations in Minkowski space-time. These equations relate geometric objects and do not need coordinates to be expressed. For example, the evolution of a particle can be described in terms of a world line $\sigma(\tau)$, and a 4-velocity $u = d\sigma(\tau)/d\tau$ that have a meaning independently of any coordinates system. The coordinate-independent proper time $\tau$ is a suitable parameter describing the evolution of physical systems in the relativistic laws of motion. Similarly, in geometric language, electromagnetic field is described by second-rank, antisymmetric tensor $F$, which also requires no coordinates for its definition. This tensor produces a 4-force on any charged particle given by $m du/d\tau = eF\cdot u$.

When coordinates are chosen, one may work with components, instead of with geometric objects. Relying on the geometric structure of Minkowski space-time, one defines the class of inertial frames and adopts a Lorentz frame with orthonormal basis vectors. Within the chosen Lorentz frame, Einstein's synchronization of distant clocks and Cartesian space coordinates are enforced. Hence, in Lorentz coordinates we have the well-known diagonal Minkowski metric tensor $g_{\mu\nu} = \mathrm{diag}(1, -1, -1, -1)$. Then, as is also well-known, any two Lorentz frames are related by a Lorentz transformation, which preserves the metric tensor components, so that the law of motion becomes

\begin{eqnarray}
&& m\frac{d^2 x_{\mu}}{d\tau^2} = e F^{\mu\nu}\frac{dx_{\nu}}{d\tau}~ ,\label{DE}
\end{eqnarray}
in any Lorentz coordinate system \cite{MTW}. It is extremely important to underline here that Einstein's synchronization of distant clocks is conventionally chosen in any Lorentz frame, when writing down Eq. (\ref{DE}) and that this choice affects the fourth component of Eq. (\ref{DE}). The importance of this observation will be clear later on.

We now consider a relativistic particle, accelerating in a given inertial frame (to fix ideas it can be the lab frame), and we analyze its evolution within the framework of special relativity. Investigation of the time dependence is best performed in the particle's rest frame where, by definition, the particle remains still. However, the problem of constructing frames associated with non-inertial motions is complicated in special relativity. If a relativistic particle accelerates in an inertial frame, its rest frame is non-inertial, and Lorentz transformations cannot be used to map observations made in the rest frame, back to the laboratory frame. To get around that difficulty, one introduces an infinite sequence of comoving frames. At each instant, one picks a  Lorentz frame  centered on the particle and moving with it. In this frame the particle is at rest for an instant. Then, an instant later, the particle's velocity changes to a new value and observation of the particle is performed with the help of  another  Lorentz frame of the sequence, centered on the particle and moving with it at the new velocity. The coordinates systems associated with the comoving sequence are assumed to be associated to an orthonormal tetrad of basis vectors. The comoving sequence of frames just described can be constructed by choosing, for each value of $\tau$ along $\sigma$, an inertial system  whose origin coincides with $\sigma(\tau)$ and whose $x'_0$-axis is tangent to $\sigma$ at $\sigma(\tau)$. Therefore, the zeroth basis vector $e'_0$ is identical to the 4-velocity $u$. In the tetrad basis $e'_i(\tau)$ the particle has proper 4-velocity $u = (c, 0, 0, 0)$ and 4-acceleration $a = (0,a_1,a_2,a_3)$ \cite{Frid,MTW,GEL}.

The next step consists in relating the basis vectors of the tetrad  $e'_0(\tau)$, $e'_1(\tau)$, $e'_2(\tau)$, $e'_3(\tau)$ at any proper time $\tau$  to the basis vectors $e_0, e_1, e_2, e_3$ of the lab frame by a Lorentz transformation
$e'_\mu(\tau) = \Lambda^\nu_\mu(\tau)e_\nu$. Therefore, the basis vectors at two successive instants must also be related to each other by a Lorentz transformation. In the lab frame one thus has a coordinate representation of the world-line as  $\sigma(\tau) = (t(\tau), x_1(\tau), x_2(\tau), x_3(\tau))$. The covariant particle trajectory $\vec{x}_{cov}(t)$ is calculated by projecting this world-line  to the lab frame basis  and using the lab time $t$ as a parameter for the trajectory curve.

We claim, and this claim is quite central for our reasoning, that there is a fundamental reason for a difference between the non-covariant particle trajectory $\vec{x}(t)$ calculated by solving corrected  Newton's equations and the covariant particle trajectory $\vec{x}_{cov}(t)$ calculated by projecting world line onto the lab frame basis as indicated above.  As just seen shown, the trajectory $\vec{x}_{cov}(t)$ is viewed from the lab frame as the result of Lorentz transformations $\Lambda^\nu_\mu(\tau)$ that depend on the proper time. As a result, the composition law that follows from the group property of the Lorentz transformation  is used to express the conditions of co-moving sequence of frames tracking a particle. In contrast to this, $\vec{x}(t)$ follows from the solution of the corrected Newton's equations and does not include Lorentz transformation composition law.

We can flesh out more the previous arguments, and discuss the difference between the trajectories $\vec{x}(t)$ and $\vec{x}_{cov}(t)$ by means of an explicit example. We will see how, depending on the way we calculate the trajectory, a composition of  Lorentz transformations will be involved (or not) yielding a Wigner rotation contribution (or not) to the trajectory seen from the lab frame. In order to do so, we first need two introductory remarks.

First, as is known, the composition of non-collinear Lorentz boosts does not results in a different boost but in a Lorentz transformation involving a boost and a spatial rotation, the Wigner rotation \cite{WI,WI1,WI2}. Suppose that our  particle moves along an arbitrary accelerated world-line. As just discussed, the basis vectors of the tetrad defining the instantaneously co-moving frames are related to the basis vectors of the lab frame  by a  Lorentz transformation depending on the proper time i.e. $e'_\mu(\tau) = \Lambda^\nu_\mu(\tau)e_\nu$. The most general Lorentz transformation  $\Lambda^\nu_\mu(\tau)$  can be  uniquely separated into a pure Lorentz boost followed by spatial rotation. Therefore, as seen from the lab frame, the space-like vectors of the tetrad (those with indexes $\mu = 1,2,3$) rotate relative to  the Cartesian axes of the lab frame. The expression for an infinitely small rotation angle  is given by \cite{Rit}

\begin{eqnarray}
	\delta \Phi = 	\left(1 - \frac{1}{\gamma} \right)\frac{\vec{v}\times d\vec{v}}{v^2}	
=	\left(1 - \frac{1}{\gamma} \right) \delta \eta ~,
	\label{d}
\end{eqnarray}
where $d\vec{v}$ is the vector of an infinitely small  velocity change due to acceleration, $\Phi$  is the Wigner rotation angle of the space-like vectors of tetrad, and $\eta$ is the orbital angle of the particle in the lab frame. From Eq. (\ref{d}) follows that in the ultra relativistic limit $\gamma \longrightarrow \infty$, space-like vectors of tetrad rotate exactly as the velocity vector $\vec{v}$.

Second, in order to analyze the finite size of a relativistic object one can extend the concept of "instantaneously comoving frame" or "infinitesiemal coordinate system" to a "local coordinate system " covering a finite domain \cite{MTW}. Consider, as before, the world-line $\sigma(\tau)$. For a fixed value of $\tau$, the three space-like basis vectors $\vec{e'}_1(\tau), \vec{e'}_2(\tau), \vec{e'}_3(\tau)$ are applied at the event $\sigma(\tau)$: these basis vectors span over a space-like hyperplane. The typical point of this hyperplane can be represented in the form $\vec{X'} = X'_1\vec{e'}_1(\tau) + X'_2 \vec{e'}_2(\tau) + X'_3\vec{e'}_3(\tau)$, where $\vec{X'}$ is the separation vector from $\sigma(\tau)$. Here the three numbers $X'_1, X'_2, X'_3$ play the role of Euclidean coordinates in the hyperplane. We then assign to each point on the spatial hyperplane the coordinates $X'_0, X'_1, X'_2, X'_3$, with $X'_0 = \tau$. These "local coordinates" approximate a Lorentz frame in the immediate neighborhood of the accelerated observer.

Having finished with the introductory remarks, we can now turn our attention to our concrete physical example on the difference between $\vec{x}(t)$ and $\vec{x}_{cov}(t)$, which is actually involving a Wigner rotation. Such effect may be of practical importance in the analysis and interpretation of experiments with ultrarelativistic modulated electron bunch at XFELs.  An ultrarelativistic electron bunch in an XFEL modulated at nm scale is a macroscopic (in the 10 $\mu$m scale) finite-size object. Therefore one needs a local coordinate system, rather than an instantaneously co-moving frame to follow the object evolution. With the help of the local coordinate system described above we can use Eq. (\ref{d}) to consider the case when the modulated ultrarelativistic electron bunch is kicked by weak dipole field. The bunch velocity is perpendicular to the wavefront of the modulation upstream of the kicker.  If the velocity of the modulated electron bunch is close to the velocity of light, Lorentz transformations work out in such a way that the rotation angle of the modulation wavefront $\delta \Phi$ coincides with the angle of rotation of the velocity $\delta \eta$. Relativistic kinematics shows the surprising effect that the orientation of the modulation wavefront is readjusted along the new direction of the electron bunch. In other words, when the evolution of the electron bunch modulation is treated according to relativistic kinematics, the orientation of the modulation wavefront in the  ultra-relativistic asymptotic is always perpendicular to the electron bunch velocity. This is plausible if one keeps in mind that the wavefront of a laser pulse behaves precisely in the same way: during the motion along a curvilinear trajectory, the wavefront of the radiation is always perpendicular to the direction of motion of the laser pulse.

In contrast, conventional  particle tracking, which is performed with the  corrected Newton's second law Eq. (\ref{N}), yields the result that a kick results in the change in the trajectory of the electron bunch, but the orientation of the modulation wavefront remains as before.  In other words, the kick results in a difference between directions of the electron motion and the normal to the wavefront. Within the lab frame, the particles motion follows the corrected Newton equations,  and there is no Wigner rotation.

This  serious discrepancy between the results "conventional" and "covariant" particle tracking naturally brings up the question: which of these results is the correct one? The answer is sophisticated. Within the framework of dynamics only, both results describe correctly the same physical reality. The expressions for the particle trajectories are different only because they are based on the use of different clock synchronization conventions. The wavefront of the modulated electron bunch can be considered  as a plane of simultaneous events. Establishing simultaneous events is only a matter of convention and the orientation of the wavefront of an ultra relativistic electron bunch has no definite objective meaning. In contrast to this, the direction of electron bunch propagation downstream the kicker obviously has a direct objective meaning and does not depends on the choice of clock synchronization.

In particular notice that while Eq. (\ref{DE}) was explicitly given together with Einstein's synchronization of distant clocks, Eq. (\ref{N}) was not. In the next sections we will demonstrate the incompatibility between  the "3+1" or "conventional" manner of performing particle trajectory calculations, based on Eq. (\ref{N}),  and the standard approach to electrodynamics, which deals with the Maxwell's equations in their usual forms.

\section{Electrodynamics. Space-time geometry}

In a coordinate-free formulation of electrodynamics, the electromagnetic field is represented by an electromagnetic tensor field $F = F(x)$ on the space-time. The source of the field is the electromagnetic current $j(x)$, which is a four-vector field.  Written in geometric language, the electrodynamics laws are

\begin{eqnarray}
&&\bigtriangledown\cdot^\ast F = 0  ~,\cr &&
\bigtriangledown\cdot F = 4\pi j~ ,\label{G}
\end{eqnarray}
where the gradient operator $\bigtriangledown$ is also four-vector and where $^\ast F$ is the dual tensor. In a Lorentz coordinate system $(^\ast F)^{\alpha\beta} = (1/2)\epsilon^{\alpha\beta\gamma\delta}F_{\gamma\delta}$ where $\epsilon^{\alpha\beta\gamma\delta}$ is the Levi-Civita pseudo-tensor \cite{MTW}.

The geometric equations Eq. (\ref{G}) are represented in the lab frame (with Lorentz coordinates on it) as the
usual Maxwell's equations Eq. (\ref{Max}). The charge and current densities Eq. (\ref{CD}) must be written as av4-vector current  by representing the charge world-line in the lab frame:

\begin{eqnarray}
x_{\mu}(\tau) = [c t(\tau), x_1(\tau), x_2(\tau), x_3(\tau)]  ~ ,\label{WL}
\end{eqnarray}
and integrating over the proper time with an appropriate additional delta function. One thus obtains \cite{D}

\begin{eqnarray}
	&& j_{\mu}(x) = ec\int d\tau  u_{\mu}(\tau) \delta^4(x - x(\tau)) ~ ,\label{CA}
\end{eqnarray}
where the charge 4-velocity $u_{\mu}(\tau)$ and the 4-vector coordinate $x_{\mu}(\tau)$ are solutions of the covariant dynamics equation Eq. (\ref{DE}). The integration over the proper time $\tau$ leads to

\begin{eqnarray}
	&& j_{\mu}(\vec{x},t) = eu_{\mu}(t)\delta^3(\vec{x} - \vec{x}_{cov}(t)) ~ .\label{CL}
\end{eqnarray}

We thus obtain

\begin{eqnarray}
	&&\rho(\vec{x}, t) =  e\delta(\vec{x} - \vec{x}_{cov}(t)) ~,\cr &&
	\vec{j}(\vec{x},t) = e\vec{v}_{cov}(t)\delta(\vec{x} - \vec{x}_{cov}(t))~, \label{CD1}
\end{eqnarray}

where $\vec{v}_{cov} = d\vec{x}_{cov}/dt$.

In the previous sections we showed that $\vec{x}_{cov}(t)$ and $\vec{x}(t)$ differ from each other. The central point of our criticism is that in all standard derivations the trajectories in the source part of the usual Maxwell's equations Eq. (\ref{Max}) were identified with the trajectories obtained in the "3+1" manner i.e. with the use of the corrected Newton's second law Eq. (\ref{N}). In other words, $\vec{x}(t)$ was always used, instead  of  $\vec{x}_{cov}(t)$ as it must be. We claim that  a solution of Maxwell's equations in their usual form based on the  results of conventional  particle tracking  $\vec{x}(t)$ cannot be used for the explanation of  experimental facts and that Maxwell's equations Eq. (\ref{Max}) are compatible, instead, with results of covariant particle tracking Eq. (\ref{CD1}).

\section{The effect of aberration of light as an example}

In this section we will consider an application of our way of coupling fields and particles in the context of the effect of aberration  of light, that is a change in the direction of light propagation ascribed to boosted sources. In particular we will expand on our previous claim by showing that the effect of aberration of light  cannot be explained on the basis of conventional coupling fields and particles.

Let us consider the case of radiation emitted by macroscopic time-varying charge and current densities. At a more fundamental level, they are just moving charges. We consider a plane full of these sources, all oscillating together, with their motion in the plane, such that they all have the same amplitude and phase.

According to the principle of relativity, usual Maxwell's equations Eq. (\ref{Max}) can always be used in any inertial frame where sources are at rest. The same considerations apply when sources are moving in non-relativistic manner. In particular, when oscillating, charged particles emit radiation, and in the non-relativistic case, when the velocities of oscillating charges $v_n \ll c$, dipole radiation will be generated and described with the help of the Maxwell's equations in their usual form,   Eq. (\ref{Max}). Here we will limit our consideration to the case of sources moving in a non-relativistic way.

Maxwell's equations can be manipulated mathematically in many ways and be casted in a form more suitable for certain applications.  For example, from Maxwell's equations we can obtain an equation which depends only on the electric field vector $\vec{E}$:

\begin{eqnarray}
c^2 \nabla^2 \vec{E} - \frac{\partial^2 \vec{E}}{\partial t^2} =
4\pi \frac{\partial^2 \vec{P}}{\partial t^2} -4\pi c^2
\vec{\nabla}\left(\vec{\nabla}\cdot \vec{P}\right) ~.
\label{nablaE2bis}
\end{eqnarray}
Here $\vec{P}$ is the dipole momentum density.

Instead of using directly the field equation in the form of Eq. (\ref{nablaE2bis}), we can use the Green's theorem to express the Fourier-transformed of Eq. (\ref{nablaE2bis}) in integral form. We first apply a temporal Fourier transformation to Eq. (\ref{nablaE2bis}) to obtain the inhomogeneous Helmholtz equation

\begin{eqnarray}
c^2 \nabla^2 \vec{\bar{E}} +\omega^2 \vec{\bar{E}}  = -4\pi \omega^2
\vec{\bar{P}} - 4\pi c^2 \vec{\nabla}\left(\vec{\nabla}\cdot
\vec{\bar{P}}\right) ~, \label{nablaE2tris}
\end{eqnarray}
where $\omega$ is the frequency and time dependence $\exp(-i\omega t)$ is understood. Note that here $\vec{\bar{E}}$ and $\vec{\bar{P}}$ are the temporal Fourier transform of $\vec{E}$ and $\vec{P}$ respectively. Here  we assume that the paraxial approximation is applicable. It is then possible to neglect the gradient term $\vec{\nabla}\left(\vec{\nabla}\cdot \vec{\bar{P}}\right)$ in the right part of the Helmholtz equation, Eq. (\ref{nablaE2tris}).

We then introduce a Green function for the Helmholtz wave equation, $G(\vec{r}, \vec{r'})$, defined as

\begin{eqnarray}
\left(\nabla^2 + \omega^2/c^2\right) G(\vec{r}, \vec{r'}) = -  \delta
\left(\vec{r} - \vec{r'}\right) ~. \label{GreenH}
\end{eqnarray}
For unbounded space, the Green's function describing outgoing waves is given by

\begin{eqnarray}
G(\vec{r}, \vec{r'}) =  \frac{1}{4\pi} \frac{\exp\left[i \omega
	|\vec{r} - \vec{r'}|/c\right]}{|\vec{r} - \vec{r'}|}  ~. \label{Grr1}
\end{eqnarray}
With the help of Eq. (\ref{Grr1}) we can write a formal solution for the field equation Eq. (\ref{nablaE2tris}) as:

\begin{eqnarray}
\vec{\bar{E}} = (\omega^2/c^2)\int d\vec{r'} ~G(\vec{r}, \vec{r'})
 \vec{\bar{P}}(\vec{r'}) ~. \label{grensol}
\end{eqnarray}
This is just a mathematical description of the concept of Huygens' secondary sources and waves, and is of course  well-known, but here we still recall that it follows directly from  the Maxwell's equations. We may consider the amplitude  of the beam radiated by our plane of oscillating charges as a whole to be the resultant of radiated spherical waves. The charges lying on the plane give rise to spherical radiated wavelets, and these combine according to  Huygens' principle to form what is effectively a radiated wave. If the plane of charges is  the $xy$-plane,  then the Huygens' construction shows that plane wavefronts will be emitted along the $z$-axis.

Now imagine that the $xy$-plane of oscillating charges is kicked along the $x$ direction in the lab frame, so that now all emitters move at constant speed $v_x$ along the  plane of sources.  After the boost along the $x$-axis, Cartesian coordinates transform as $x' = x - v_xt$, $y' = y$, $z' = z$. This transformation is completed with the invariance in simultaneity  $\Delta t = \Delta t'$. The absolute character of the temporal coincidence of two events is a consequence of the absolute concept of time $t' = t$ in Newtonian kinematics. Within the lab frame there is an Euclidean three-space, a global (absolute) time $t$, and the Newton's second law of motion. We come to a situation when there is a uniform motion of the elementary sources along the plane of simultaneity with velocity $v_x$. In other words, all sources in the plane have the same phase of oscillation. If we now apply  Maxwell's equations Eq. (\ref{Max})  to the case of kicked sources, the Huygens' construction shows  that  plane wavefronts are emitted along the $z$-axis, without a change in the direction of emission, i.e, without light aberration.  We come to the conclusion that the effect of aberration of light cannot be explained on the basis of the conventional "3+1" coupling of usual Maxwell's equations Eq. (\ref{Max}) with charge trajectories  measured  within the lab frame.

\subsection{Einstein's synchronization convention}

The explanation of the effect of aberration of light  presented in well-known textbooks is actually based on the use of  relativistic kinematics i.e. on a transformation from the lab inertial frame to another moving inertial frame. These two different inertial frames with Lorentz coordinates are related to one another in a non-Newtonian way.  In this  approach,  quantities involved do not pertain the "3+1" space and time of the laboratory frame, but are rather quantities in 4D Minkowski space-time.

To see this we consider a Lorentz boost in the $x$-direction to describe the uniform translation along the $x$-axis in the laboratory frame. Transformation of observations in the lab frame $S$ to the moving frame $S'$ can be made by a Lorentz transformation. On the one hand, it is well known that the wave equation (and also Huygens' principle ) remains  form-invariant with respect to Lorentz transformations. On the other hand, if make a Lorentz boost, we automatically introduce a time transformation  $t' = \gamma (t - xv_x/c^2)$ and the effect of this transformation is just a rotation of the radiation wave front in the lab frame. This is because the effect of this time transformation is just a dislocation in the  timing of processes, and  has the effect of rotating  the plane of simultaneity on the angle  $v_x/c$ in the first order approximation. In that case, due to uniform motion along the $x$ direction, the elementary sources produce a spatial phase modulation, i.e. a  chirp, given by $\omega x v_x/c^2$, where $\omega$ is the oscillation frequency. It is not hard to see that this chirp is simply a trivial consequence  of the relativity of simultaneity between the two Lorentz frames $S$ and $S'$. As a consequence of this linear phase chirp, the radiation wavefront rotates in the lab frame. In fact, the sources in the plane $xy$  have a spatially-dependent  chirp of phase oscillation. Then,  Huygens' construction shows that a linear phase chirp  implies radiation of a plane wave with the angle  $\theta = v_x/c$ from the $z$-direction, which is the usual aberration of light effect.

It should be clear that in "3+1" space and time  where we live as observers, we cannot associate phase of oscillating charges  with the result of a Lorentz transformation. Within the lab frame, the spatial phase modulation   $\omega x v_x/c^2$   has only formal significance. This means that, for instance, the measurement of oscillating phases  within the lab frame will give the usual  result of a uniform phase along the $xy$ plane. Within the lab frame, the motion of the particles looks precisely the same as predicted by Newtonian kinematics and the phase chirp  does not exist at all.

The difference between the results obtained by using the Lorentz transformation between $S$ and $S'$, predicting a phase chirp,  and the result of an actual  measurement within the lab frame, where no chirp is observed, brings up the question: which of these results is correct? The answer is baffling: within the framework of dynamics only, both results correctly describe the same physical reality. Different expressions for phase distribution along the plane of emitters are the result of the use of different clock synchronization conventions in the lab frame. We cannot give any experimental method by which the phase shift between two  particles in oscillating in different places can be ascertained: determining distant simultaneity is only matter of convention.

We have thus arrived at an explanation of the effect of aberration of light by using relativistic transformations. Let us summarize the result of standard  derivations, which  deal with the relation between two Lorentz frames, i.e. inertial frames with specific Lorentz system of space-time coordinates. We must use a Lorentz boost (i.e. Einstein's convention of clock synchronization in both inertial frames, as prescribed by the use of Eq. (\ref{DE})), because we want to solve the electrodynamics problem based on the  usual Maxwell's equations in the case of moving sources. Maxwell's equations in their usual form  can certainty be used in any inertial frame  for a source at rest, and the transformation connecting two inertial frames with Lorentz coordinates is a Lorentz transformation. Then, since the usual Maxwell's equations are preserved by Lorentz transformations, usual Maxwell's equations hold in the lab frame as well. However, source trajectories measured (or calculated by conventional particle tracking) in the lab frame obey Eq. (\ref{N}) and pertain a different choice of distant clock synchronization: as such, they are not compatible with  Maxwell's equations in their usual form. The usual Maxwell's equations should be solved with the source trajectory which is viewed from the lab frame as a result of Lorentz transformation connecting the lab frame with the moving frame where source is at rest. This corresponds to the choice of Einstein synchronization convention in both frames and in particular, in the lab frame.

\subsection{Absolute time convention}

It should be noted that there is another satisfying way of coupling fields and particles, which consists in using  charged particles trajectories measured within the lab frame.  As we already discussed above, usual Maxwell's equations cannot be employed, because we want to solve electrodynamics problem based on the use quantities from "3+1" space and time. Within the lab frame, as shown before,  we have elementary sources in uniform motion  along the $xy$ plane of simultaneity with velocity $v_x$. One can explain the effect of aberration of light in the "3+1" space and time similarly as done before, i.e. modeling the kick as a transformation from the inertial lab frame to the moving inertial where emitters are at rest. These two inertial frames are related to one another by a Galilean transformation. After the Galileo boost along the $x$-axis, Cartesian coordinates and time transform as

\begin{eqnarray}
x' = x - v_xt, ~ y' = y, ~ z' = z, ~ t' = t.  ~, \label{GT}
\end{eqnarray}
The Galilean time appears to be an absolute time, since it is the same in all inertial frames.  The uniform phase along the plane $xy$ is now viewed from the lab frame as a result of the Galilean transformation connecting the lab frame with the moving frame where emitters are at rest. We have thus obtained the following result: when a Galilean transformation is chosen, the sources in the moving frame viewed from the lab frame move exactly as one can see by performing a direct measurement in the lab frame.

We would now like to discuss how Galilean transformations can be understood in terms of the theory of relativity. Since the inception of special relativity, most researchers assume that Lorentz transformations directly follow from the postulates of the theory of relativity. However, these postulates alone are not sufficient to obtain Lorentz transformations: one additionally needs to synchronize spatially separated moving clocks with the help of light signals. If this is done by using the Einstein's synchronization convention, then  Lorentz transformations follow. However, if the same clocks are synchronized following a different synchronization convention, other transformations follow. In order to get a Galilean transformation,  we should synchronize clocks in the source rest frame.  Next, in order to perform measurements (or calculations) in the lab frame, it is necessary to synchronize the  clocks at rest in the lab frame where the source is in uniform translational motion along the $x$ axis.  This can be done with the help of the moving clocks, simply by adjusting clocks at rest to zero whenever a moving clock that shows zero  flies past a clock at rest (see e.g. \cite{GU}).

In agreement with the principle of relativity, usual Maxwell's equations can always be exploited in a moving inertial frame where emitters are at rest. However, the transformation connecting two inertial frames with absolute time synchronization is a Galilean transformation, and  Maxwell's equations do not remain form-invariant with respect to Galilean transformations. In fact, the d'Alambertian in the left part of Eq. (\ref{nablaE2bis}) is not a Galilean invariant. The change of  Maxwell's equations under Galilean transformations can be found from the knowledge of the metric tensor components $g_{\mu\nu}$. Due to tensorial character of $g_{\mu\nu}$, starting from the diagonal form in the rest frame and applying a Galilean transformation we obtain $g_{00} = 1-v_x^2/c^2$, $g_{01} = v_x/c$, $ g_{11} = g_{22} = g_{33} = -1$. As a result, in the "3+1" space and time we have much more complicated electrodynamics of moving charges than usual. Let us consider the Galilean transformation of the d'Alembertian. We want to know   how Eq. (\ref{nablaE2bis}), which describes the radiation process in the moving frame where emitters are at rest, appears from the point of view of the lab frame with absolute time synchronization.

In the moving frame fields are expressed as a function of independent variables $x', y', z'$, and $t'$. Now, the variables $x',y',z',t'$ can be expressed in terms of the independent variables $x, y, z, t$ by means of Galilean transformation, so that fields can be expressed in terms of  $x, y, z, t$. From the Galilean transformation Eq. (\ref{GT}) after partial differentiation, one obtains

\begin{eqnarray}
\frac{\partial }{\partial t} = \frac{\partial }{\partial t'} - v_x \frac{\partial }{\partial x'}~ , ~
\frac{\partial }{\partial x} = \frac{\partial }{\partial x'} ~.
\end{eqnarray}
Hence the  d'Alembertian $\Box'^2 = \nabla'^2 - \partial^2/\partial(ct')^2$ transforms into

\begin{eqnarray}
&& \Box^2 = \left(1-\frac{v_x^2}{c^2}\right)\frac{\partial^2}{\partial x^2}  - 2\left(\frac{v_x}{c}\right)\frac{\partial^2}{\partial t\partial x}
+ \frac{\partial^2}{\partial y^2} + \frac{\partial^2}{\partial z^2}
- \frac{1}{c^2}\frac{\partial^2}{\partial t^2} ~
\end{eqnarray}
where coordinates and time are transformed according to a Galilean transformation. After properly transforming the d'Alembertian we can see that the inhomogeneous wave equation for the electric field in the lab frame has nearly but not quite  the usual, standard form that takes when there is no common, uniform translation of charges in the transverse direction with velocity $v_x$. The main difference consists in the "interference" term $\partial^2/\partial t\partial x$ which arises from the non-diagonal component of the metric tensor $g_{01} = v_x/c$. To get around this difficulty, we observe that this equation can always be simplified. The trick needed here is to make a change of the time variable. In order to the eliminate the interference term we make a variable transformation $t' = t - x v_x/c^2$ . In the new variables, i.e. after the Galilean coordinate transformation and time shift, we obtain the  d'Alambertian in the following form

\begin{eqnarray}
\Box^2 =	\left(1-\frac{v_x^2}{c^2}\right)\frac{\partial^2}{\partial x^2} + \frac{\partial^2}{\partial y^2} + \frac{\partial^2}{\partial z^2}
	- \left(1-\frac{v_x^2}{c^2}\right)\frac{1}{c^2}\frac{\partial^2}{\partial t^2} ~.
\end{eqnarray}
A change in the scale of time and coordinate along the direction of uniform motion in the ratio $\gamma$ leads to the usual Maxwell's equations. In particular, the  d'Alambertian  $\Box'^2 = \nabla'^2 - \partial^2/\partial(ct')^2$  transforms into  $\Box^2 = \nabla^2 - \partial^2/\partial(ct)^2$  when coordinates and time are transformed according to Galilean transformation followed by the  variable changes specified above. The overall combination of Galileo transform and variable changes actually yields to Lorentz transformation in "3+1" space and time

\begin{eqnarray}
x' = \gamma(x -  v_x t), ~ y' = y, ~ z' = z, ~ t' = \gamma(t - x v_x/c^2)  ~, \label{LT}
\end{eqnarray}
but in the present context they are only to be understood as a useful mathematical device that allows one to solve the electrodynamic problem in the "3+1" space and time with minimal effort. Obviously, transforming to new variables leads to the usual Maxwell's equations. Time shift results in a slope of the simultaneity plane and sources in the $xy$ plane are characterized by a chirp of phase oscillation. Then, the  radiation field is given by Eq. (\ref{nablaE2bis}), and plane waves are radiated at an  angle $\theta = v_x/c$, yielding the phenomenon of light aberration: the two approaches, treated according to Einstein's or absolute time synchronization conventions give the same result. In other words, the Galilean transformation completed by the introduction of the new variables  is mathematically equivalent to the Lorentz transformation described in the previous subsection.

We state that the variable changes performed above have no intrinsic meaning - their meaning only being assigned by convention. In particular, one can see the connection between time shift $t' = t - xv_x/c^2$ and the issue of clock synchrony. A change in the scale of time  also is unrecognizable from physical viewpoint. It does not matter which convention and hence transformation or "translation"  is used to describe the same reality. What matters is that, once fixed, such convention should be applied and kept in a consistent way.

\section{Experimental test}

We showed that the results of conventional theory of radiation by moving charges are not consistent with the principle of relativity. There are experimental results that actually confirm the conclusion that the conventional ("3+1") particle tracking  must be replaced by a covariant 4D space-time particle tracking according to our previous description, so that the result of the tracking be consistent with the usual Maxwell's equations. For example, in the preceding sections it was pointed out that even a well-established as the effect of aberration of light cannot be explained  on the basis of the conventional coupling between fields and particles. The experimental checks of the effect of aberration of light  are perhaps the most direct and  convincing evidence that covariant coupling of fields and particles is correct, and does not need modification. However, there are other experiments that can be considered and that we will discuss in this section.

The arguments of Section 3 show that for relativistic motion results of covariant particle tracking differ from results of conventional particle tracking in the case of particles accelerating along  a curved trajectory.

At present,  relativistic synchrotron-cyclotron radiation results are textbook examples and do not require a detailed description. In the ultra relativistic limit the particle emissivity is provided by well-known analytical formulas, which represent the spectral and angular behavior of synchrotron-cyclotron radiation emitted by an electron moving in a constant magnetic field, and having an ultra relativistic velocity component perpendicular to it. However, here we must emphasize  that a correction of conventional synchrotron-cyclotron radiation theory is expected in the light of the pointed difference between conventional and covariant particle tracking.

Our criticism of all standard derivations is that one cannot perform, as always done, identification of the trajectories in the source part of the usual Maxwell's equations with the trajectories measured (or calculated) in the "3+1" space and time. One way to demonstrate incompatibility between the standard approach to electrodynamics, which deals with the usual Maxwell's equations and particle trajectories measured within the lab frame it is to make a direct laboratory test of the synchrotron-cyclotron radiation theory. In other words, we are stating here that, despite the many measurements done during decades, synchrotron-cyclotron radiation theory is \textit{not} an experimentally well-confirmed theory. The most elementary of the effects that represents a crucial test of the coupling fields and particles, is in fact a red shift of the critical wavelength of synchrotron-cyclotron radiation,  which is direct consequence of the difference between Einstein's velocity addition law  and the law of addition of velocities  in Newtonian kinematics. Such a measurement is critical, in the sense that it can confirm or falsify our theory on the coupling between fields and particles, and has never been performed to our knowledge.

The dynamical evolution in the lab frame described by Eq. (\ref{N}) is based on the use of the lab frame time $t$ as independent variable. In this case the trajectory $\vec{x}(t)$ can be seen, from the lab frame view, as the result of successive Galileo boosts that track the motion of the accelerated particle. The usual Galileo rule for addition of velocities is used to fix the  Galileo boosts tracking a particular particle along its motion.

In contrast, according to covariant particle tracking, any transformation of observations in the comoving inertial frame, where a particle is instantaneously at rest, to the lab frame can be done by Lorentz transformations. If the relativistic particle is accelerated in the lab frame, one can think of successive Lorentz transformations tracking  its motion. In this case, it is the Einstein's rule for addition of velocities, which  is used to fix the Lorentz transformations tracking that particular particle along its motion.

In \cite{OURS2} we focused on the description of an experiment that can reveal the difference between the predictions of conventional synchrotron radiation theory and our correction in some commonly used setup. We considered the simple case when an ultrarelativistic electron beam is kicked by a weak dipole field before entering a downstream undulator and we studied the process of emission of spontaneous undulator radiation with and without transverse kick. According to conventional particle tracking, after the beam is kicked there is a trajectory change, while the electron velocity remains as before. The prediction of the conventional theory is that if an electron beam is at perfect undulator resonance without kick, then after the kick the same electron beam must be at perfect resonance in the kicked direction. This is plausible, if one keeps in mind that, after the kick, the particle have the same velocity and emit radiation in the kicked direction owing to the Doppler effect.

However, covariant particle tracking clearly shows that the electron after the kick has lower velocity due to Einstein's rule of addition of velocities for non collinear boosts \footnote{We stess that, in agreement with what we said up to now, this is the velocity that has to be inserted into Maxwell's equations in their usual form, not the velocity measured by an observer in the lab frame.} . It follows, in terms of corrected synchrotron radiation theory, that the electron emits spontaneous undulator radiation with a red shift in the kicked direction. Since, in terms of the conventional approach, the same particle would emit spontaneous undulator radiation without red shift in the kicked direction, we insist on performing a not complicated and not expensive experiment at 3rd generation light sources that can confute the conventional theoretical approach.

Synchrotron radiation from bending magnets is emitted within a wide range of frequencies. The possibility of using narrow bandwidth sources in an experimental study on the difference between the predictions of conventional synchrotron radiation theory and our correction looks more attractive. This allows one to increase the sensitivity of the output intensity on the red shift, and to relax the requirement on the beam kicker strength and photon beam line aperture.

Undulators as sources of quasi-monochromatic synchrotron radiation produce light in sufficiently narrow bandwidth. They cause the electron beam to follow a periodic undulating trajectory with the consequence that interference effects occur. Undulators have typically many periods. The interference of radiation produced in different periods results in a bandwidth that scales as the inverse number of periods. Therefore, the use of insertion devices installed at third generation synchrotron radiation facilities would allow us to realize a straightforward increase in the sensitivity to the red shift at a relatively small kick angle, $\theta < 1/\gamma$.

In \cite{OURS2} we also stressed that the presence of red shift in undulator radiation automatically implies  the same effect in conventional cyclotron radiation theory. In fact, the conventional theory predicts that there should be no red shift for  radiation emitted by an electron  which has velocity directed  along and across the magnetic lines of force. In the ultrarelativistic limit, the particle emissivity is provided by well-known analytical formulas, which represent the spectral and angular behavior of cyclotron radiation emitted by an electron moving in a constant magnetic field, having a non-relativistic velocity component parallel to the field, and an ultrarelativistic velocity component perpendicular to it. According to the conventional approach, exactly as for the undulator case, the angular-spectral distribution of radiation is a function of the total velocity of the particle due, again, to the Doppler effect. We note that cyclotron-synchrotron radiation emission is one of the most important processes in plasma physics and astrophysics and the results of an experimental test of conventional radiation theory  would constitute a truly critical experimental test for a much wider part of  physics than that of synchrotron radiation or X-ray free electron laser sources.

There is another interesting problem where our correction of synchrotron radiation theory is required, which involves the production of coherent undulator radiation. Let us consider a microbunched ultrarelativistic electron beam kicked by a weak dipole field before entering a downstream undulator. We want to study the process of emission of coherent undulator radiation from such setup. According to conventional particle tracking, based on Eq. (\ref{N}),  after the beam is kicked there is a trajectory change, while the orientation of the microbunching phase front remains as before. In other words, the kick results in a difference between the direction of the electron motion and the normal to the phase front. In standard electrodynamics, coherent radiation is emitted in the direction normal to the microbunching wavefront. Therefore, according to the conventional coupling of fields and particles, when the angular kick exceeds the divergence of the output coherent radiation, emission in the direction of the electron beam motion is strongly suppressed. In Section 3 we have shown that our manifestly covariant coupling of fields and particles predicts an effect in complete contrast to the conventional treatment. Namely, in the ultrarelativistic limit, the plane of simultaneity, that is wavefront orientation of the microbunching,  is always perpendicular to  the electron beam velocity. As a result, we predict strong emission of coherent undulator radiation  from the  modulated electron beam in the kicked direction.

From a pragmatic viewpoint, physical theories should be able to predict experimental results in agreement with measurements, i.e. they should "work". The fact that our theory  predicts reality in a satisfactory way is well-illustrated by comparing the prediction we just made with the results of an experiment involving "X-ray beam splitting" of a circularly-polarized XEL pulse from the linearly-polarized XFEL background pulse, a technique used in order to maximize the degree of circular polarization at XFELs\cite{NUHN}.

Circularly polarized X-ray radiation is needed for investigating magnetic materials and for dealing with other important material science issues. Therefore, the production of X-ray radiation with a high degree of circular polarization is strongly pursued at XFEL facilities. The quality of the output radiation is optimized in the case  of a full-length helical undulator, but this option remains unrealized at the first stage of many XFEL projects, due to technical challenges and costs related with the production of long helical undulators. The choice of a relatively short helical undulator constitutes a reasonable compromise. The electron beam is microbunched in the preceding planar undulator segments before being sent into a short helical radiator. This enhances the power of the circularly polarized pulse by several orders of magnitude compared with spontaneous emission. The short helical undulator is then said to be operating in "afterburner configuration". Using an afterburner is in fact an economically and technically convenient method to generate circularly polarized radiation. However, as an unavoidable by-product, the linearly-polarized background radiation  from the main undulator is superimposed with the circularly polarized pulse.

An ultimate  solution to reduce this background component was proposed in \cite{L}. After the main planar undulator the electron beam is deflected by a bending system  and subsequently sent through the helical afterburner. If the  microbunching structure can be preserved,   the helical undulator still produces intense coherent radiation, but now  linearly- and circularly-polarized radiation are well-separated  by the bending system. The challenge here consists in the preservation of the microbunching on the scale of the radiation wavelength, which is of order of one nanometer, as the electron beam progresses through the bending system. Note that in order to effectively separate the linearly and the circularly polarized radiation components, the bending angle must be large compared to the divergence of the coherent radiation.

Also consider that according to the conventional theory, the microbunching is preserved after passing through a simple dipole, but it keeps its original direction. Then, as discussed above, coherent radiation emission is exponentially suppressed, unless the microbunching is rotated by the bending angle, so that it becomes orthogonal to the new direction of propagation. In  order to do that,  a significant effort is required.  The  isochronous  bending system  proposed in \cite{L} for the European XFEL requires about 87 m long of total length (which is compatible with the space available in one of the European XFEL tunnels). It consists of 33 magnets, including 8 dipoles, 9 quadrupoles, and 16 sextupoles. The design in \cite{L} is based on the use of conventional particle tracking and  XFEL codes.

In reality, no isochronous bending system was actually needed at the LCLS facility to achieve intense emission of coherent, highly circularly polarized radiation. In its baseline mode of operation, the LCLS  generates intense X-ray pulses from a planar undulator.  A 3.2 m-long compact helical undulator segment was installed in place of the last LCLS undulator segment \cite{NUHN}, and used in  afterburner mode. The efficiency  of this mode of operation was tested, and a maximum contrast ratio between circularly and linearly polarized radiation components of about ten was reported in \cite{NUHN}. The same reference further reports a final improvement of the contrast ratio by "X-ray beam splitting". This was achieved by kicking the electron beam before entering the helical undulator, so that electron beam and background radiation could pass through the helical undulator at different angles. A single corrector magnet, that is a weak dipole magnet, placed at the end of the  last planar undulator section was used to control the magnitude of the kick. Full elimination of the linear polarized component was achieved through spatial separation combined with transverse collimation.  In was confirmed that the degree of circular polarization was very close to $100 \% $\cite{NUHN}. Experimental results  in \cite{NUHN} clearly show an additional red shift of the resonance wavelength in the kicked direction:  the maximum power of the coherent radiation after the kick was reached when the undulator was detuned to be resonant to a lower longitudinal velocity after the kick.

The "X-ray beam splitting" experiment at the LCLS \cite{NUHN}  apparently demonstrated that after a microbunched electron beam is kicked on a large angle compared to the divergence of the FEL radiation  \footnote{ The tuning limit of deflection angle  was set at $\sim$ 5 rms of FEL radiation divergence by beamline aperture, see Fig. 14 in \cite{NUHN}}, the microbunching wavefront is readjusted along the new direction of motion of the kicked beam. This is the only way to justify coherent radiation emission from the undulator placed after the kicker and along the kicked direction.

Summarizing, the authors of \cite{NUHN} found that coherent undulator radiation was produced in the kicked direction and red-shifted compared to the nominal electron beam energy and undulator strength. These results came unexpectedly, but from a practical standpoint, the  "apparent wavefront readjusting"  immediately led to the realization that the unwanted, linearly-polarized radiation background could be fully eliminated without extra-hardware. In other words a single corrector,  already part of the baseline installations in the intersection between undulator segments, effectively worked as the complex and expensive bending system designed according to the theory of conventional particle tracking in \cite{L}. The results of the "beam splitting" experiment at the LCLS demonstrated that even the direction of emission of coherent undulator radiation is beyond the predictive power of the conventional theory.

In this paper we showed that the authors of \cite{NUHN} actually witnessed an apparent wavefront readjusting due to the phenomenon of Wigner rotation, but they never drew this conclusion. We are indeed first in considering the idea that results of the conventional theory of radiation by relativistically moving charges are not consistent with the principle of relativity. In previous literature, identification of the trajectories in the source part of the usual Maxwell's equations with the "true" trajectories measured in the lab frame has always been considered obvious. The impact of \cite{NUHN} on our studies was immediate. At first their result was quite mysterious, but we now understand why it should be so. We analyzed the situation when a microbunched beam is kicked, and we discovered that several strange phenomena should occur. Now everything fits together, and our theory, albeit shocking, shows the existence of both wavefront readjusting and  red shift of the resonance wavelength \footnote{It is necessary to mention that in the case of the beam splitting experiment at the LCLS we deal indeed with an ultra relativistic  electron beam with ($c-v\simeq 10^{-8}c )$, and with a transverse velocity after the kick which is very much smaller than the speed of light ($(v_x/c)^2 \ll 10^{-8}$), so that the theoretical studies presented in \cite{OURS1,OURS2,OURS3,OURS4}  yield a correct quantitative description of the beam splitting experiment at the LCLS and, in particular, of the red-shift of the resonance wavelength in the kicked direction.}. The theory of relativity as a theory of 4D space-time with pseudo-Euclidean geometry has had more than hundred years of history and development, and rather suddenly it has begun to be fully exploited in practical  ways in accelerator physics.

\section{Discussion and Conclusions}

Eq. (\ref{N}), Eq. (\ref{Max}) and Eq. (\ref{CD}) form the basis of conventional coupling of fields and particles in accelerator and plasma physics, which is considered for more than hundred years as relativistically correct. In textbooks on  electrodynamics it is argued  that the properties of radiation by moving relativistic charges can be calculated by using only the lab frame without reference to Lorentz transformations.

In this paper we presented our arguments, based on special relativity, against this conventional way of coupling fields and particles. We found that, upon a choice of convention, the velocity of a relativistic particle in the lab frame can be defined in two ways.  We called one of them, $d\vec{x}(t)/dt$, the "3+1" velocity. This is obtained by a direct measurement in the lab frame (or by conventional particle tracking calculations). The other is the more mathematical, more abstract velocity $d\vec{x}_{cov}(t)/dt$, which we  call "covariant velocity". This velocity is obtained by projecting the particle world-line to the lab frame basis, where the lab frame is assumed to be a Lorentz frame. It turns out that these velocities differ from one another and that in the theory of electrodynamics of moving charges it is the covariant velocity, and not  the "3+1" velocity which is connected with the usual Maxwell's equations.

The theory of relativity shows that the intuitive concept of particle trajectory  based on direct measurements in the lab frame cannot be used in all situations. While this statement seems in disagreement with the idea that measurement is actually defining a physical quantity, let us take as an example the relation between the "true" measured velocity $d\vec{x}(t)/dt$ and the "covariant" velocity $d\vec{x}_{cov}(t)/dt$. The "true" velocity $d\vec{x}(t)/dt$ is the result of a direct measurement within the lab frame. The quantity $d\vec{x}_{cov}(t)/dt$ differs from the "true" velocity and has only formal significance. However, we are better off using covariant trajectories when we want to solve the electrodynamics problem based on Maxwell's equations in their usual form. In fact, the use of "true" trajectories implies the use of much more complicated field equations.

A comparison with classical non-relativistic physics might help to make our point.
In classical mechanics one can consider two momenta (see, for example \cite{J,F}). One of them is the "kinematic momentum". This is the momentum obtained by multiplying mass by velocity, $\vec{p} = m\vec{v}$. It can be obtained by a direct measurement within the lab frame. The other is a more mathematical, more abstract momentum, called the "dynamical momentum", $\vec{P}$. The two definitions differ by the vector potential, since $\vec{P} = m\vec{v} + e\vec{A}$. It turns out that in classical mechanics it is the dynamical $\vec{P}$ momentum that is connected with standard canonical dynamical equations. In other words, if we want to solve the dynamics problem based on the  usual canonical equations of motion, we must use the dynamical momentum. We might use the kinematic momentum, but we would have much more complicated equations of motion. To get around this difficulty, we observe that these equations can always be simplified. The trick needed here is to make a change of variables. Transforming to new variables $\vec{P}$ leads to the usual canonical equations. However, the new dynamical momentum has only formal significance in the sense that a trajectory measurement in the lab frame will always give back the usual kinematic momentum $m\vec{v}$.

Let us now return to the Galilean transformations. It is generally believed that Galilean transformations are incorrect because they do not preserve the form-invariance of Maxwell's equations under a change of inertial frame.
To quote e.g. Rosser \cite{R}: "Now if the co-ordinates and time are transformed using the Galilean transformations, Maxwell's equations do not obey the principle of relativity. [...] If the Galilean transformations were assumed to be correct, Maxwell's equations could only be hold in the one reference frame and should have been possible to identify this absolute system by electrical and optical experiments. However in practice it proved impossible to identify by means of experiments any absolute reference frame for the laws of optics and electromagnetism. [...] This meant abandoning the Galilean transformations in favor of the Lorentz transformations."

This idea is a part of the material in well known textbooks (see e.g. \cite{F}). As a result many physicists still tend to think of Galilean transformations as old, incorrect transformations between spatial coordinates and time. We disagree with this point of view. The special theory of relativity is the theory of four-dimensional space-time with pseudo-Euclidean  geometry. From this viewpoint, the principle of relativity is a simple consequence of the space-time geometry, and the space-time continuum can be described in arbitrary coordinates. In the process of "translation" to arbitrary coordinates, the geometry of the four-dimensional space-time does not change.  Therefore, contrary to the view presented in textbooks,  Galilean transformations are actually compatible with the principle of relativity although, of course, they alter the form of Maxwell's equations.

A comparison with the three-dimensional Euclidean space  might help here. In the usual 3D Euclidean space, one can consider a Cartesian coordinate system ($x,y,z$), a cylindrical coordinate system ($r,\phi, z$), a spherical coordinate system ($\rho, \theta, \phi$), or any other. Depending on the choice of the coordinate system one respectively has  $ds^2 = dx^2 + dy^2 + dz^2$, $ds^2 = dr^2 + r^2d\phi^2 + dz^2$, or $ds^2 = d\rho^2 + \rho^2d\theta^2 + \rho^2\sin \theta^2d\phi^2$, where $ds$ is the distance between two closely separated points. The metric actually does not change, but the components of the metric do, depending only on the choice of coordinates.
In general, in fact, we write $ds^2 = g_{ik}dx^idx^k$. Considering Cartesian coordinates we will always have $g_{ij} = \mathrm{diag}(1,1,1)$. Similarly, Lorentz transformations between inertial frames with Einstein coordinates leave the components of the metric tensor unvaried.

Let us now discuss an example, that is  characteristic  for the traditional use of the theory of relativity in the explanation of electromagnetic phenomena from the point of view of two moving inertial frames. This example appears in many standard thextbooks. To quote Feynman, Leiton and Sands \cite{F}: "It is interesting to discuss what it means that we replace the old  transformation between the coordinates and time with the new one, because the old one (Galilean) seems to be self-evident, and the new one (Lorentz) looks peculiar. We wish to know whether it is logically and experimentally possible that the new, and not the old, transformation can be correct. [...] When this (i.e. Galilean transformation of Maxwell's equations) is tried,the new terms that had to be put into the (Maxwell's) equations led to predictions of new electrical phenomena that did not exist at all when tested experimentally, so this attempt had to be abandoned.".

Our point of view is in disagreement with this. The "true" trajectory of a charge $\vec{x}(t)$, which is found by direct measurement in the lab frame, or calculated by using the corrected Newton's second law, is viewed from the lab frame as the result of successive Galileo transformations. In other words, within the lab frame, the motion of the particles looks precisely the same as predicted by Newtonian kinematics, and relativistic effects like Wigner rotations do not exist. This is due to a particular choice of synchronization convention in the lab frame, which we called the "absolute time convention". In agreement with the principle of relativity, usual Maxwell's equations can be used in a moving inertial frame where the charge is instantaneously at rest. However, the transformation connecting any comoving frame to the lab frame in the case of the absolute time convention is a Galilean transformation, and Maxwell's equations do not remain invariant with respect to Galilean transformation. When a Galilean transformation of Maxwell's equations is tried, the new terms that have to be put into the Maxwell's equations lead to predictions of radiation phenomena  that actually do  exist  when tested experimentally. To be specific, the main extra-term consists in the "interference" term (like $\partial^2/\partial t\partial x$) in the d'Alambertian, which arises from the non diagonal component of the metric tensor. These new terms, introduced  in the inhomogeneous wave equation, give the effect of light aberration  and explain coherent radiation emission of a microbunched, ultra-relativistic beam in the kicked direction. Alternatively, one may use the trajectory $\vec{x}_{cov}(t)$, which is compatible with the usual Maxwell's equations and is found by manifestly covariant dynamical equations. This is viewed from the lab frame as the result of successive Lorentz transformations. The Lorentz transformation composition law is used to express the conditions of comoving sequence of frames tracking a particle. In this case, a Wigner rotation arises and explains light aberration and coherent emission. The two approaches give, in fact,  the same results, and it does not matter which  transformation (Galilean or Lorentz) is used: they both describe the same reality.

\section{Acknowledgements}

We greatly thank Martin Dohlus  and Igor Zagorodnov for useful discussions.

\end{document}